\documentclass[prl,preprint,superscriptaddress,preprintnumbers,amsmath,amssymb,floats]{revtex4}
\pdfoutput=1
\usepackage{txfonts}
\usepackage{amssymb}
\usepackage{graphicx}
\usepackage{sidecap}
\usepackage{color}

\begin{document}

\title{Local antiferromagnetic exchange and collaborative Fermi surface as key ingredients of high temperature superconductors}
\author{Jiangping Hu}\email{jphu@iphy.ac.cn }  \affiliation{Beijing National Laboratory for Condensed Matter Physics, and Institute of
Physics, Chinese Academy of Sciences, Beijing 100190, China}
\affiliation{Department of Physics, Purdue University, West Lafayette, Indiana 47907, USA}
 \author{Hong Ding}\email{dingh@iphy.ac.cn}   \affiliation{Beijing National Laboratory for Condensed Matter Physics, and Institute of
Physics, Chinese Academy of Sciences, Beijing 100190, China}

\begin{abstract}
\textbf{Cuprates, ferropnictides and ferrochalcogenides are three 
%Cuprates \cite{bed}, ferropnictides \cite{Hosono} and ferrochalcogenides \cite{FeTe,ChenXL} are three 
classes of unconventional high-temperature superconductors, who share similar phase diagrams in which superconductivity develops after a magnetic order is suppressed, suggesting  a strong interplay between superconductivity and magnetism, although the exact picture of this interplay remains elusive.  Here we show that there is a direct bridge connecting antiferromagnetic exchange interactions determined in the parent compounds of these materials to the superconducting gap functions observed in the corresponding superconducting materials.  High superconducting transition temperature is achieved when the Fermi surface topology matches the form factor of the pairing symmetry favored by local magnetic exchange interactions. 
%Pairing symmetry is also selected simultaneously by the matching.  
Our result offers a principle guide to search for new high temperature superconductors.
\\
}
\end{abstract}

\maketitle

In a conventional superconductor, superconductivity emerges from a normal metallic state below a critical transition temperature $T_c$.  This phase transition can be pictured well in the reciprocal space where each point labels electron momentum.  In the normal state, electrons occupy the reciprocal space to form a Fermi sea.  In the superconducting state, a pair of electrons with opposite momenta near the surface of the Fermi sea are bound together to form a Cooper pair by an attractive force generated through absorption and emission of phonons - the vibrations of  crystal lattice. Superconductivity is globally formed when all pairs move coherently.  The pairing strength can be determined by measuring an energy gap, $\Delta$,  opened on the Fermi surface.   In a standard BCS superconductor \cite{BCS}, $T_c$  is proportional to the gap value at zero temperature: $\Delta = 1.57k_BT_c$.   Within this traditional picture of superconductivity, magnetism is considered to be an enemy of  superconductivity because it  breaks Cooper pairs.   Furthermore, if the phases of Cooper pairs change signs in the reciprocal space, even non-magnetic impurities are harmful to superconductivity \cite{anderson}.

In contrast,  the three known classes of high-$T_c$ superconductors (cuprates, ferropnictides and ferrochalcogenides) apparently violate many of these conventional wisdoms mentioned above \cite{hightcreview, john}.  First, the superconductivity in these high-$T_c$ materials develops from a `bad metal' state whose resistivity is several orders of magnitude higher than those of normal metallic states in conventional superconductors.  Second, strong magnetism is involved in the `bad metal' parent state and superconductivity occurs when long-range magnetic order is suppressed, as shown in Fig.~1 where the phase diagrams of ferropnictides and cuprates are plotted.  Note that magnetic orders may even coexist with superconductivity in both ferropnictides and electron-doped cuprates. Third, the ratio between superconducting gap and critical transition temperature is much larger than the BCS ratio of 1.57 \cite{gapcuprates, hding,Zhou_gap,Nakayama, Wang_122Se,Zhang_122Se,Mou_122Se}.  Finally, superconductivity in high-$T_c$ superconductors is rather robust against impurities \cite{balatsky,john}, contrary to conventional superconductors.  Among all these peculiarities, the superconductivity living at the edge of magnetic orders is still the most intriguing phenomenon.  

It is true that even today,  there is still a lack of  comprehensive understanding how superconductivity arises by doping parent compounds and suppressing magnetic orders.   Even in a simple model, such as, the well-known $t-J$ model proposed for cuprates \cite{anderson2,zhangrice},  the physics  is too rich and complex to be understood in a fully controllable manner.  Nevertheless,  even if a complete picture of high-$T_c$ superconductors has yet been reached,   earlier investigations of the $t-J$ model for cuprates \cite{kotliar,AtoZ}  have  provided some deep insights about the interplay between magnetism and superconductivity:  (i) carrier doping can destroy long-range magnetic orders, but not  short-range magnetic exchange interactions;  (ii)  short-range magnetic exchange interactions  can be responsible for pairing electrons and drive superconductivity.   These two insights emphasize closely bound singlet electron pairs in real space  rather than in momentum or reciprocal space.   The attraction force is generated by local instantaneous  antiferromagnetic (AF) exchange interactions, differing from the retarded attractive force  generated through emitting and absorbing ``gluon" in conventional superconductors.      If one closely examines the meanfield solution of $t-J$ model \cite{kotliar},  there are two additional implications related to reciprocal space properties that are also critical to high $T_c$:  (iii)  superconductivity develops if the hopping parameter $ t $ (or band width)  can be renormalized such that  they become comparable to magnetic exchange interactions; (iv) high $T_c$ and selection of pairing symmetry  are simultaneously obtained if  the reciprocal form factor of the pairing symmetry provided by local magnetic exchange interactions takes large absolute value on the Fermi surface. These two implications, especially (iv), were not manifestly emphasized before in addressing how high $T_c$  can be achieved.  

 Strictly speaking, the $t-J$ model which includes no double occupancy constraint is a model for strongly correlated electron systems. Iron-based superconductors is more itinerant than cuprates. Between ferropnictides and ferrochalcogenides, the former is also considered to more itinerant.  Neverthless,  disregarding many microscopic electronic differences among the three classes of high-$T_c$ superconductors, based on the spirit of the above four points, here we show that there exists a basic paradigm to unifiedly understand both cuprates and iron-based superconductors, including ferropnictides and ferrochalcogenides:  the key ingredients in the determination of high $T_c$ and pairing symmetries are local AF exchange interactions in real space and Fermi surface topology  in reciprocal space that matches to the pairing form factor provided by the AF interactions.   Such a paradigm will help to predict new high-$T_c$ superconductors and provide a guide to modify the properties of a material to  increase $T_c$.  \\
\\
{\bf Effective magnetic exchange interactions}

First, we examine the magnetic exchange interactions of parent compounds of high-$T_c$ superconductors. In all three classes of high-$T_c$  materials, the transition metal atoms form a tetragonal square lattice.  Their parent compounds  exhibit  distinct magnetically ordered states \cite{caf,bcaf,bcaf2,baf}. 
 In Fig.~2, we illustrate their magnetic exchange interactions and ordered spin configurations.   
 
In cuprates, the magnetic order is a  checkerboard AF state with an ordered wavevector $(\pi,\pi)$  as shown in Fig.~2a.  This state can be naturally derived from a Heisenberg model  where only the nearest neighbor (NN) AF interaction $J_1$  is important and longer range  magnetic exchange interactions can be ignored. Microscopically, $J_1$ is generated by the superexchange mechanism mediated through oxygen atoms located in the middle of  two NN copper atoms. 

In ferropnictides,  the magnetic order is a collinear AF (CAF) state with an ordered wavevector $(\pi,0)$ \cite{caf} as shown in Fig.~2b.   This magnetic state can be obtained in a $J_1-J_2$ Heisenberg model with  $J_1< 2J_2$  \cite{local1,local2,local3,it3,local4},  where $J_2$ is the $2_{nd}$ NN magnetic exchange interaction. 
 The measurement of spin wave excitations in the parent compounds of ferropnictides indicates that  both  $J_1$ and $ J_2 $ are AF \cite{zhao1}.    

In the 11-ferrochalcogenide, FeTe,   the magnetic order is a bi-collinear AF (BCAF) state with an ordered wavevector $(\pm \frac{\pi}{2},\pm\frac{\pi}{2})$ \cite{bcaf,bcaf2} as shown in Fig.~2c.  To obtain this magnetic state, a  third NN ( $3_{rd}$ NN)  AF exchange coupling $J_3$ is needed \cite{ma,cfang, hu4}. In fact, the analysis of spin wave excitations in FeTe shows that  a ferromagnetic (FM)  $J_1$  and an AF $J_3$  must be included while $J_2$ does not differ significantly from  ferropnictides \cite{lip2}.  The magnetic exchange interactions are confirmed again in the 122-ferrochalcogenide,  K$_{0.8}$Fe$_{1.6}$Se$_2$, which exhibits a block AF state with an ordered wavevector $(\frac{3\pi}{5},\frac{\pi}{5})$ \cite{baf}.   Analyzing spin wave excitations in the block-AF state yields similar magnetic exchange interactions  as FeTe \cite{my}. 

Table 1 summarizes important AF exchange interactions in five different high-$T_c$ superconductors.
It is worth to note that in both statically ordered CAF and BCAF phases of  ferropnictides and ferrochalcogenides, the spin wave excitations  suggest that  $J_1$ must have different values on links with different spin configurations.
This difference can be explained if a NN biquadratic effective spin interaction \cite{local4,hu4}  is included. However,  since such a term does not play a role in providing superconducting pairing, we will not discuss it further.\\
\\
{\bf Reciprocal form factors of pairing symmetries and determination of high $T_c$}

 Second, we examine the possible pairing symmetries and their reciprocal form factors determined from the corresponding magnetic exchange interactions. 
For  an $s$-wave and $d$-wave spin singlet pairing superconductor,  only AF exchange interactions play a role in pairing electrons.    The explicit pairing forms determined from the AF magnetic models discussed above for five different high-$T_c$ materials are listed in Table 1 and their detailed derivation is explained in the supplementary material.

Finally, after knowing the  form factors of possible pairing symmetries , one can apply the standard Eliashberg equation to determine the pairing symmetry and the transition temperature. The (iii) and (iv) implications really stem from this step.  Taking a one-band system with a single AF magnetic exchange interaction as an example,  $T_c $ is determined by the following self-consistent meanfield equation (the generalized Eliashberg equation) \cite{kotliar} as
\begin{eqnarray}
2T_c=J_{\alpha}\sum_k |f_{\alpha}(k)|^2 g(x(k,T_c))\nonumber
\end{eqnarray}  where  $g(x)=\frac{tanh(x)}{x}$ and  $x(k,T_c)=\frac{\epsilon(k)-\mu}{2T_c} $.  $\epsilon(k)$ is the band dispersion and $f_{\alpha}(k)$ is  the corresponding pairing form factor determined by the AF exchange interaction $J_{\alpha}$. 
The function $g(x)$ is always positive and has its maximum value on Fermi surfaces. In order to obtain nonvanishing $T_c$ in the Eliashberg equation,  the band dispersion $\epsilon(k)$ has to be strongly renormalized  so  that $J_{\alpha}$ is  comparable to  the band width. In the meanfield solution of the $t-J$ model \cite{kotliar}, the non-double occupancy constraint is transferred to  strong band renormalization.   Iron-based superconductors are multi-band systems.  Similar meanfield treatment of an extended $\tilde t-\tilde J$ model has been studied in refs.\cite{seo2008,Fang2011}, where the parameters of the band structure $\tilde t$ are  presumably taken to be strongly renormalized such that they become comparable to magnetic exchange interactions.   These studies also show that  the pairing symmetry and the transition temperature are  mainly determined by the weight of the form factors near Fermi surfaces.   

Here, rather than performing calculation within a theoretical model, we take band structures and Fermi surfaces of high-$T_c$ superconductors measured by angle-resolved photoemission spectroscopy (ARPES) and calculate the overlap between the pairing form factors and the Fermi surfaces, $\sum_k|f(k)|^2\delta(\epsilon(k)-\mu)$, which is the value of the quantity on the right side of the Eliashberg equation at zero temperature that approximately determines $T_c$.   The quantitative results of the overlap in five typical high-$T_c$ superconductors are summarized in Table 1, and the detailed formula to evaluate the overlap is explained in the supplementary material. One can visualize this overlap by plotting Fermi surface and gap function in the same reciprocal space, as shown in Fig.~3. 
%Clearly, a large overlap is realized in all of high-$T_c$ superconductors.

To demonstrate the importance of this overlap in achieving high $T_c$, we illustrate the details  of Fermi surface and superconducting gap of the three classes of high-$T_c$ superconductors determined by ARPES: (i)  In Fig.~4a, we show a typical Fermi surface of cuprates (the Fermi surface of optimally doped Bi$_2$Sr$_2$CaCu$_2$O$_{8+x}$ measured by ARPES \cite{gapcuprates} is shown). In this case, it is clear from Table 1 and Fig.~3 that the $d$-wave form,  $cosk_x-cosk_y$, has a much larger overlap with the Fermi surface than the $s$-wave form. Therefore, a $d$-wave pairing symmetry with the form $cosk_x-cosk_y$ is favored.   Indeed, ARPES results  strongly support the $d$-wave form, as shown in Figs.~4d, 4g \cite{gapcuprates}. (ii)  In Fig.~4b, we show the Fermi surfaces of  ferropnictides featuring  pockets located at the $\Gamma$, M and Z points in the unfolded Brillouin zone (the Fermi surfaces of optimally hole doped Ba$_{0.6}$K$_{0.4}$Fe$_2$As$_2 $ measured by ARPES \cite{hding,Nakayama} are shown here). There are two hole pockets at $\Gamma$,  one hole pocket at Z and one electron pocket at M.   In this case, it is also clear from Table 1 and Fig.~3 that the $s$-wave  form factor  $cosk_xcosk_y$ provided by the 2$_{nd}$ NN AF $J_2$ has the maximum overlap with the Fermi surfaces.  Consequently, in a doping region where electron and hole pockets are reasonably balanced,  an $s$-wave with a symmetry form $cosk_xcosk_y$ should dominate in the superconducting state, which has  also been observed by ARPES, as shown in Figs.~4e, 4h \cite{hding,Nakayama}.  However, with a high percentage of hole or electron doping, which destroys the balance between electron and hole pockets, the AF NN $J_1$ can start to take effect on the pairing symmetry.  For example, in the case of heavily hole-doped systems where the Fermi surfaces are dominated by the hole FS pockets at $\Gamma$ (Z), the $d$-wave form  $cosk_x-cos k_y$ can strongly compete with the $s$-wave form $cosk_xcosk_y$. Indeed, there are strong experimental evidence for gap nodes in the heavily hole-doped superconductor KFe$_2$As$_2$ ($T_c\sim 3K$) \cite{Li,hashimoto}. Such a competition will weaken superconductivity as shown in refs.\cite{seo2008,Fang2011}.  (iii)  In Fig.~4c, we plot the Fermi surfaces of ferrochalcogenides for FeTe$_{0.55}$Se$_{0.45}$ \cite{ding11}, where one hole pocket at Z and one electron pocket at M are observed.  In this case, the electron pocket dominates over hole pockets.  The $s$-wave symmetry $cosk_xcosk_y$  still has a good overlap with Fermi surfaces. However, unlike the case of ferropnictides, here the NN interaction  $J_1$ is FM  so that there is no competition from the $d$-wave form $cosk_x-cosk_y$. Thus, we still expect a dominant $s$-wave pairing.  The presence of a significant $3_{rd}$ AF $J_3$ adds interesting effect on the gap function.  For an $s$-wave,  an AF $J_3$ provides an additional pairing form, $cos2k_x+cos2k_y$, which takes large values at both hole and electron pockets as well.  However, unlike  $cosk_xcosk_y$  which takes opposite sign between $\Gamma$(Z) and M,  the form  takes the same sign at $\Gamma$(Z) and M.  Therefore if we mix these two forms together and consider that the electron pockets dominates over the hole pockets,  we naturally expect that the pairing form in these materials should be proportional to $cosk_xcosk_y-\delta (cos2k_x+cos2k_y)$   with $\delta$ being positive. This pairing form exactly describes what is observed in FeTe$_{0.55}$Se$_{0.45}$, as shown in Figs.~4f, 4i \cite{ding11}. With this form, the gap on the electron pocket at M should be larger than the gap on the hole pocket at Z, as shown in Fig.~4f \cite{ding11}.  The same  analysis  can also  be applied to the recently discovered high-$T_c$ superconductor KFe$_{1.7}$Se$_2$, which only has electron pockets at M \cite{Wang_122Se,Zhang_122Se, Mou_122Se,Qian}.    With both $J_2$ and $J_3$ being AF,  the absence of hole pockets allows the gap function in the electron pockets to take large values to achieve high $T_c$. 
\\
\\
 {\bf Predictions of possible high temperature superconductors}  
 
The paradigm established here allows us to predict possible magnetic interactions and Fermi surfaces in undiscovered high-$T_c$ superconductors.  It is clear that the presence of strong local AF interactions is necessary.  Assuming that these interactions are known, we can discuss the possible matching Fermi surfaces which can lead to high-$T_c$ superconductivity in several common lattice structures. In Figs.~5a, 5b, we draw two possible Fermi surfaces that can lead to high $T_c$ for a tetragonal lattice structure. The Fermi surface in Fig.~5a  leads to an $s$-wave superconductor for a strong NN AF interaction while the one in Fig.~5b leads to a $d$-wave superconductor for a strong $2_{nd}$ NN AF interaction.  In Figs.~5c, 5d, we draw two Fermi surfaces  that can lead to $s$-wave pairing symmetry in a honeycomb lattice when the NN and $2_{nd}$ NN AF exchange interactions dominate respectively.  The detailed reciprocal pairing forms are given in the supplementary material.  The prediction for a triangle lattice with NN AF exchange interactions and $s$-wave pairing symmetry is similar to Fig.~5c with a rotation of 30 degrees of all Fermi surface around the center $\Gamma$ point.  We do not address  $d$-wave paring symmetry  in a honeycomb lattice here because the $d$-wave superconducting  state  will most likely break the time-reversal symmetry. \\
 \\
  {\bf Discussions and conclusions} 
  
  The paradigm described here is still a phenomenological,  or at most, a semi-microscopic understanding of high-$T_c$ superconductors. Nevertheless, it is already a powerful guide to understand many  unconventional properties in these materials.
  
The paradigm  suggests that the effect of electron-electron correlations is very important to high $T_c$.  It can strengthen local AF exchange interactions as well as cause strong renormalization of band structures. However, a strict Mott-insulating state is not a necessity of high $T_c$.  The AF exchange interactions  rely  more sensibly  on the electronic properties of the atoms which mediate superexchange interactions, such as oxygen in cuprates and As or Se(Te) in iron-based superconductors, rather than on-site interaction $U$.   The fact that the Eliashberg equation is still meaningful in understanding high-$T_c$ superconductivity suggests the importance of reciprocal space properties.

The paradigm also suggests that the sign change of superconducting orders in reciprocal space is not due to the positive Josephson couplings between  different Fermi surfaces. Instead, it is determined together by local AF exchange interactions and  Fermi surface topology, namely  the sign change behavior is  a derivative product,
 rather than an origin to cause high $T_c$ in the first step that  has been proposed in many weak coupling theories \cite{mazin,Kuroki,WangF}.  Since the pairing generated by the AF exchange interactions already avoids strong onsite repulsive interactions, it is inevitable that the superconducting order  has a sign-distribution in reciprocal space.  Of course,   
 Josephson couplings derived in weak coupling theories and superconducting pairing provided by local AF exchange interactions can collaborate with each other to drive higher $T_c$.  Such as in ferropnictides, the collaboration can happen if there are positive Josephson couplings between the hole pockets at $\Gamma$(Z) and electron pockets at M.  However,  in KFe$_2$Se$_2$, due to the absence of hole pockets
 at $\Gamma$(Z),  the positive Josephson couplings between two electron pockets at M will damage superconductivity if  it is $s$-wave pairing.  A verification of $s$-wave pairing symmetry in KFe$_2$Se$_2$ will be an important support for the paradigm since the positive Josephson coupling between two electron pockets results in a $d$-wave pairing symmetry \cite{maitikorsh}.

The paradigm further provides qualitative explanations of many unusual behaviors, including strong pairing, short coherence length and impurity insensitivity. The strong pairing  and the short coherence length result from instantaneous  and  short-range  attractive force generated by AF exchange 
interactions.  The superconducting states in all high-$T_c$ superconductors are rather robust against impurities, in contrast to  a conventional superconductor where even non-magnetic impurities  can significantly alter the
pairing interaction and break Cooper pairs if the superconducting order has a sign variation in the reciprocal space. Within the paradigm, the pairing force is determined rather locally and the sign change is  due to its form factor derived from local AF exchange. Therefore, if the local AF exchange interactions are not significantly altered by impurities, the pairing force is stably maintained. Although a quantitative comparison is still difficult, the effect of impurities on superconductivity should be much weaker in high-$T_c$ superconductors than in conventional superconductors.

A hidden assumption of the paradigm is that the pairing force can be smoothly derived from the local AF exchange interactions existed in the magnetic parents, which suggests that the leading AF exchange interactions should not be  drastically modified  in  doped materials.  This  assumption can be tested directly by measuring  high-energy spin excitations or other spin properties in doped compounds. In cuprates, recent experiments  using resonant inelastic
x-ray scattering have reported that many superconductors, encompassing underdoped
YBa$_2$Cu$_4$O$_8$  and overdoped YBa$_2$Cu$_3$O$_7$, exhibits damped spin excitations (paramagnons) with
dispersions and spectral weights similar to those of magnons in undoped cuprates \cite{tacon}.  In ferropnictides,  similar results have been obtained in the study of BaFe$_{2-x}$Ni$_x$As$_2$ by neutron scattering experiments \cite{pcdai}. Moreover,  it has  also been shown that the AF $J_2$ in Li$_{1-x}$FeAs \cite{lifeas},  which is already self-doped, is  similar to other parent compounds \cite{zhao1,lip2,my}.  In ferrochalcogenides,  there is a  rather robust incommensurate magnetic excitation  in all superconducting FeTe$_{1-x}$Se$_x$ samples \cite{lum,Argyriou,LeeSH}, which  suggests  $J_3$ is rather robust.

In summary, we have demonstrated that the superconducting gap symmetry and amplitude of cuprate, ferropnictide, and ferrochalcogenide high-$T_c$ superconductors, as observed by ARPES, can be naturally determined by the local antiferromagnetic exchange interactions of their magnetic parent compounds collaborating with the Fermi surface topology in the superconducting offspring compounds.  By identifying local AF exchange and collaborative Fermi surfaces as key ingredients of high-$T_c$ superconductors, we are able to predict magnetic configuration, Fermi surface topology and pairing symmetry of several undiscovered high-$T_c$ superconductors.  We believe that this phenomenological description establishes a foundation towards a microscopic theory of unconventional high-temperature superconductivity.

\textbf{Acknowledgement: } {We thank P. Richard for many valuable suggestions for the manuscript and H. Miao for assistance in constructing figures. We acknowledge B. A. Bernevig,  P.-C. Dai, D.-L. Feng, S. Kivelson, D.-H. Lee, H.-H. Wen,  X.-G. Wen, Z.-Y. Weng, T. Xiang and C.-K. Xu for valuable discussions. This work is supported by National Basic Research (973) Program of China (grants No. 2010CB92300) and Chinese Academy of Sciences (grant No. 2010Y1JB6). }

\begin{center}
\begin{table*}
 \begin{tabular}{|l|c|c|c|c|c|}\hline
AF couplings \& gap form& Bi$_2$Sr$_2$CaCu$_2$O$_{8+x}$ &Pr$_{1-x}$Ce$_x$CuO$_4$  & Ba$_{0.6}$K$_{0.4}$Fe$_{2}$As$_2$ & FeTe$_{0.55}$Se$_{0.45}$ & KFe$_{1.7}$Se$_2$ \\
 \hline
$J_1$: $s$-wave  $(cosk_x+cos k_y)/2$ &   0.03  & 0.01&  0.43 &  (0.29)  & (0.01)   \\
\hline
$J_1$: $d$-wave $(cosk_x-cosk_y)/2$ & {\color{red}0.61}& {\color{red}0.40}& 0.36 & (0.55)&(0.74)   \\
\hline
$J_2$: $s$-wave  $cosk_xcos k_y$ &   -- &  --&{\color{red}0.62} & {\color{red}0.71}    & {\color{red} 0.55}\\
\hline
$J_2$: $d$-wave $sink_xsink_y$ & -- & --& 0.03 & 0.01     & 0.05 \\
\hline
$J_3$: $s$-wave  $(cos2k_x+cos 2k_y)/2$ &   -- & -- &-- & {\color{red} 0.52}  &  {\color{red}0.31} \\
\hline
$J_3$: $d$-wave $ (cos2k_x-cos 2k_y)/2$ & -- & -- &--& 0.07& 0.11\\
\hline
 \end{tabular}
\caption{ Summary of AF exchange interactions, possible  reciprocal symmetry forms, and strength of their overlap with Fermi surfaces (shown as numbers in the table) in five different high-$T_c$ superconductors. The numbers with red color indicate the primary superconducting pairings in the corresponding materials.  The numbers with parentheses  are just for comparison since the corresponding magnetic exchange is FM.   The overlap in the electron doped cuprate Pr$_{1-x}$Ce$_x$CuO$_4$ is calculated from the band structure measured in ref.\cite{pcco}, showing a smaller value than the one obtained in hole-doped cuprates.}
 \end{table*}
 \end{center}
 
%   \begin{center}
%  \begin{tabular}{cccc}\hline
% magnetic    &Cuprates	& Ferropnictides & Ferrochalcogenides  \\
% \hline
%$J_1$ &   AF  &  AF & FM      \\
%$J_2$ &  N/A & AF & AF      \\
%$J_3$ & N/A & N/A & AF  \\       
%s-wave pairing & $\Delta_1(cosk_x+cos k_y)$ &  $\Delta_2 cosk_xcos k_y+ \Delta_1 (cosk_x+cosk_y)$   & $\Delta_2 cosk_xcos k_y+ \Delta_3 (cos2k_x+cos2k_y) $   \\
%d-wave pairing &  $\Delta_1(cosk_x-cos k_y)$ &  $\Delta_2 sink_xsin k_y+ \Delta_1 (cosk_x-cosk_y)$   & $\Delta_2 sin_xsin k_y+ \Delta_3 (cos2k_x-cos2k_y) $\\
% \hline \label{parameters}
% \end{tabular}
% \end{center}
%  	 	 

\begin{figure*}
\includegraphics[scale=0.4]{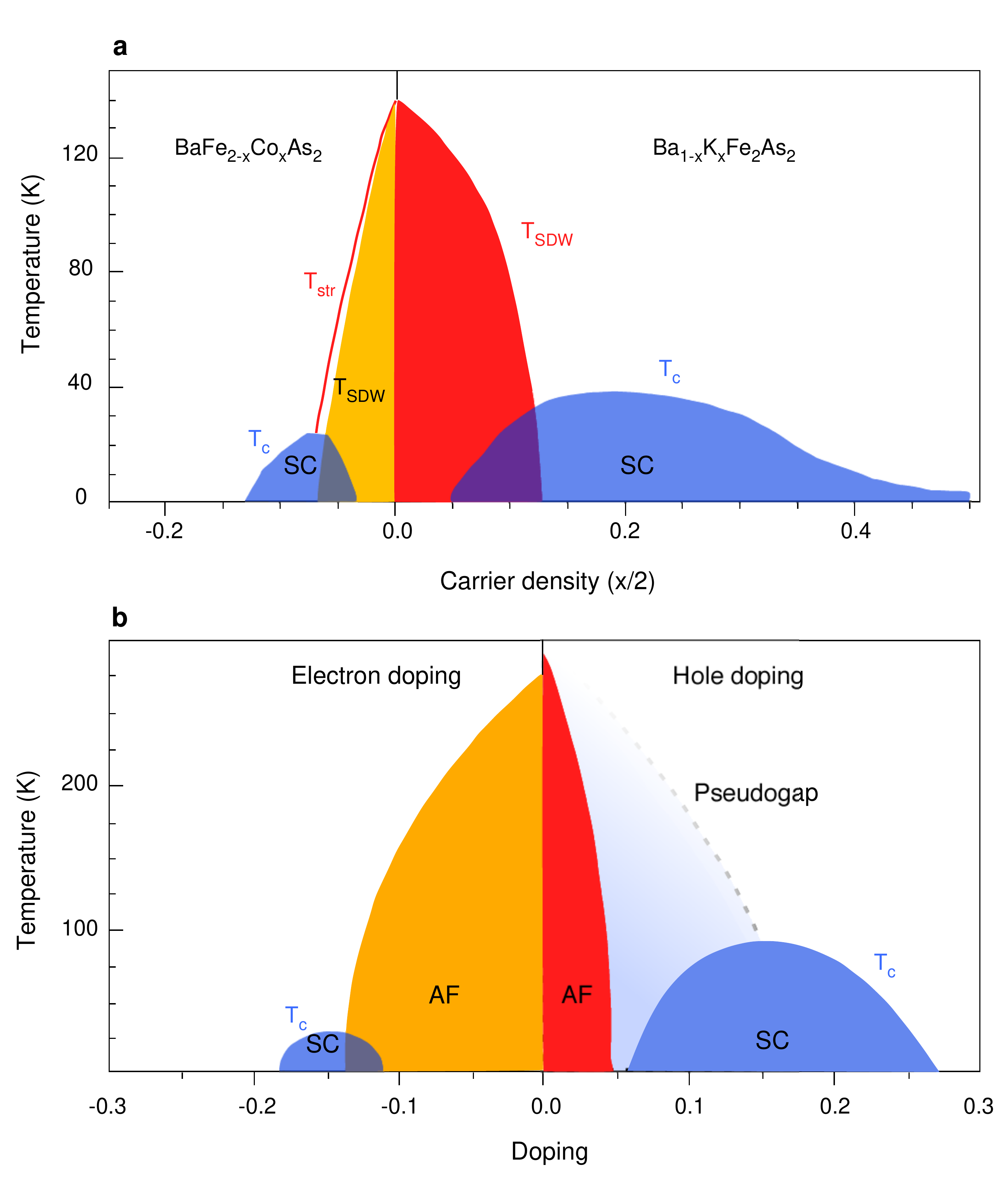}
\caption{  {\bf Phase diagrams of high-$T_c$ superconductors.}
 {\bf a,} Phase diagram of 122-ferropnictides (adopted from ref.\cite{neupane}).  {\bf b,} a generic phase diagram of cuprates. \label{fig1} }
\end{figure*}

\begin{figure*}
\includegraphics[scale=0.4]{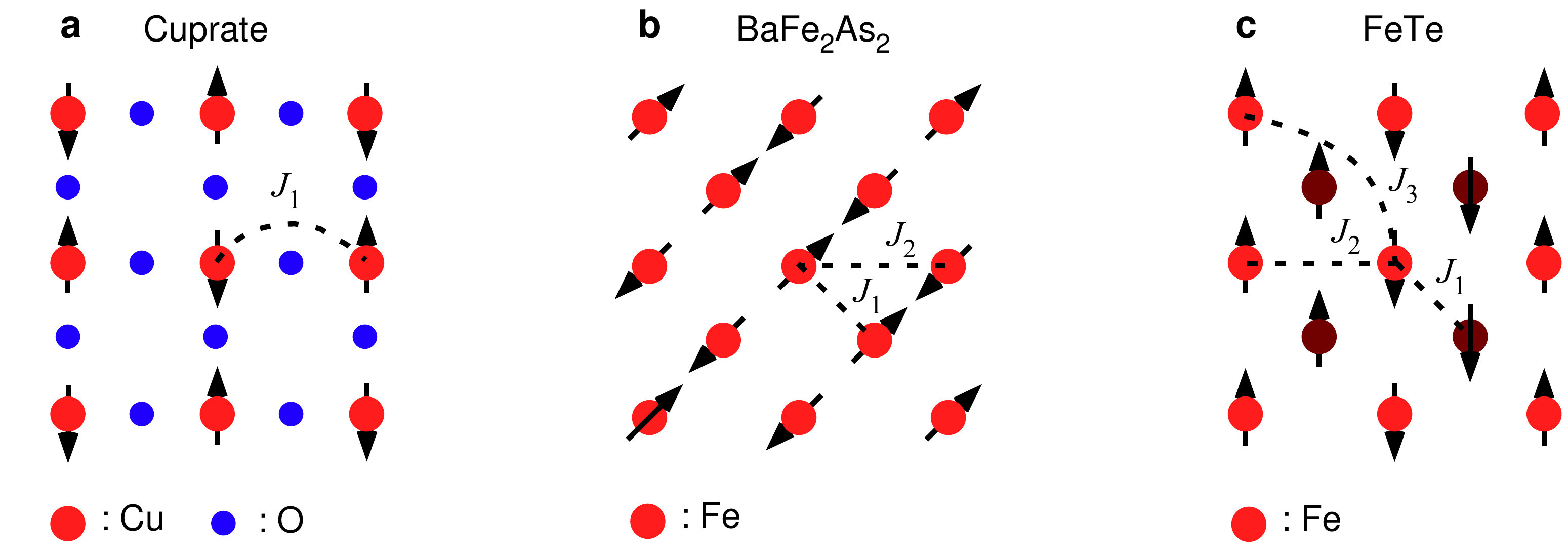}
\caption{  {\bf Magnetically ordered states of high-$T_c$ superconductors.}
 {\bf a,} checkerboard AF ordering in cuprates.  {\bf b,} collinear AF ordering in ferropnictides. {\bf c,}  bi-collinear AF ordering in ferropnictides. \label{fig1} }
\end{figure*}

\begin{figure*}
\includegraphics[scale=0.5]{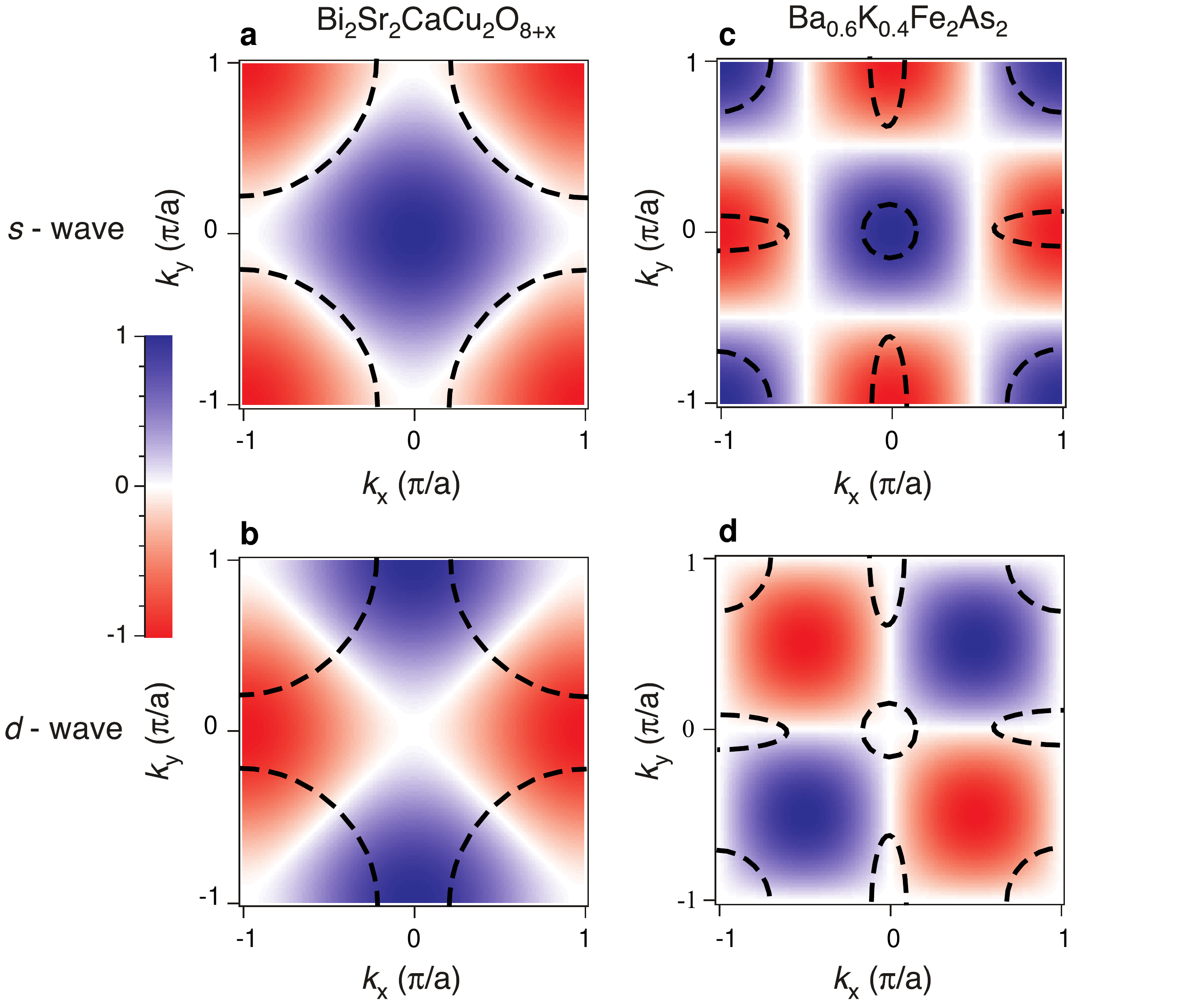}
\caption{  {\bf Visualization of the overlap between Fermi surface and gap functions.}
 {\bf a,} $d$-wave $cosk_x-cosk_y$ for optimally doped cuprate Bi$_2$Sr$_2$CaCu$_2$O$_{8+x}$.
 {\bf b,} $d$-wave $sin_xsink_y$ for optimally doped ferropnictide Ba$_{0.6}$K$_{0.4}$Fe$_{2}$As$_2$.
 {\bf c,} $s$-wave $cosk_x+cosk_y$ for Bi$_2$Sr$_2$CaCu$_2$O$_{8+x}$.
 {\bf d,} $s$-wave $cosk_xcosk_y$ for Ba$_{0.6}$K$_{0.4}$Fe$_{2}$As$_2$.
 The color bar indicates the values of the superconducting order parameters. \label{fig3} }
\end{figure*}

\begin{figure*}
\includegraphics[scale=0.4]{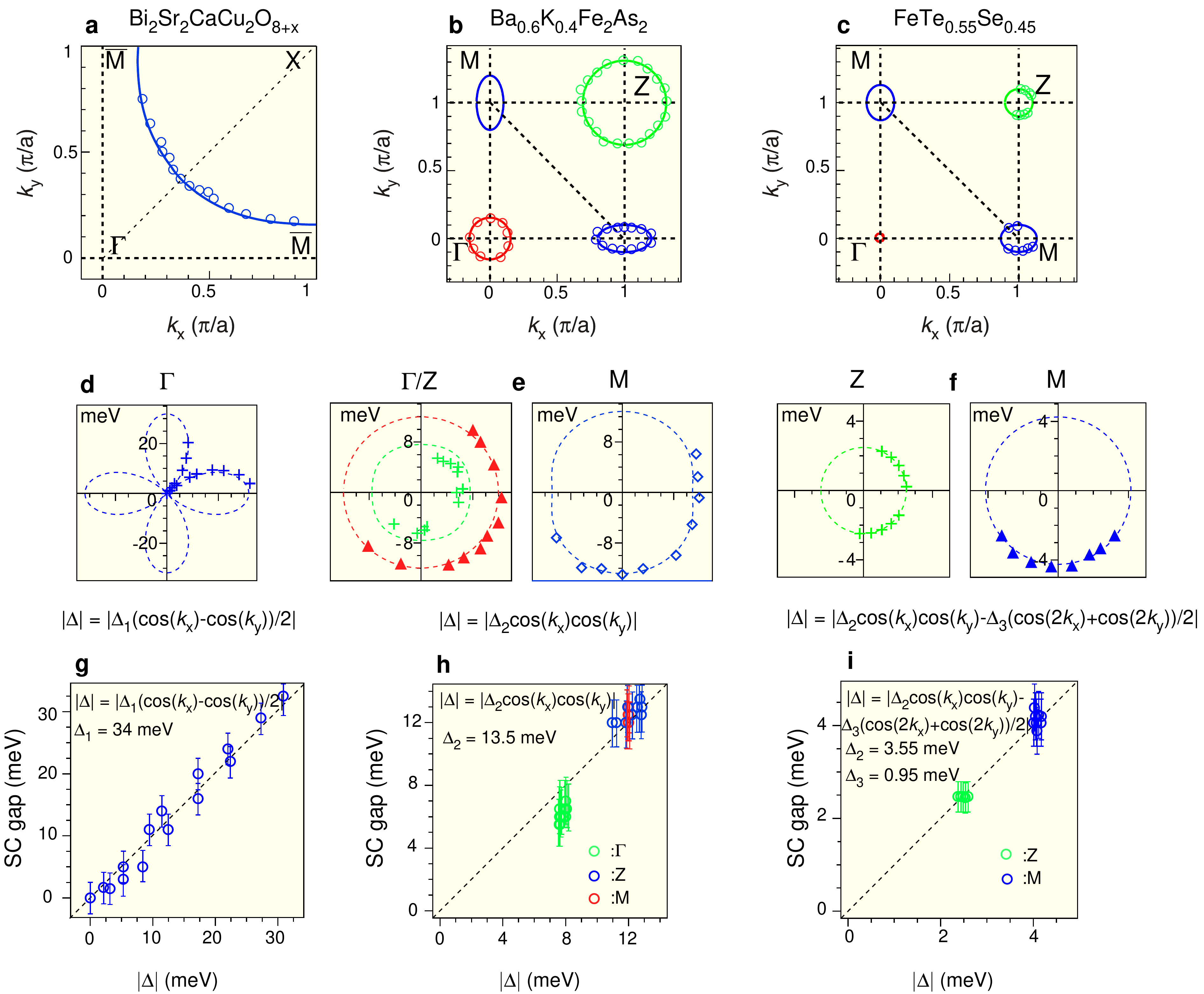}
\caption{ {\bf ARPES results of Fermi surface and superconducting gap of high-$T_c$ superconductors.}
Fermi surface topologies ({\bf a} - {\bf c}), momentum dependence of the superconducting gap in polar plots ({\bf d} - {\bf f}) (dashed lines are the corresponding gap functions plotted in the panels below), and their fits to  reciprocal symmetry forms ({\bf g} - {\bf i}) of three high-$T_c$ superconductors: Bi$_2$Sr$_2$CaCu$_2$O$_{8+x}$, Ba$_{0.6}$K$_{0.4}$Fe$_{2}$As$_2$, and FeTe$_{0.55}$Se$_{0.45}$, respectively.  \label{fig4} }
\end{figure*}

\begin{figure*}
\includegraphics[scale=0.3]{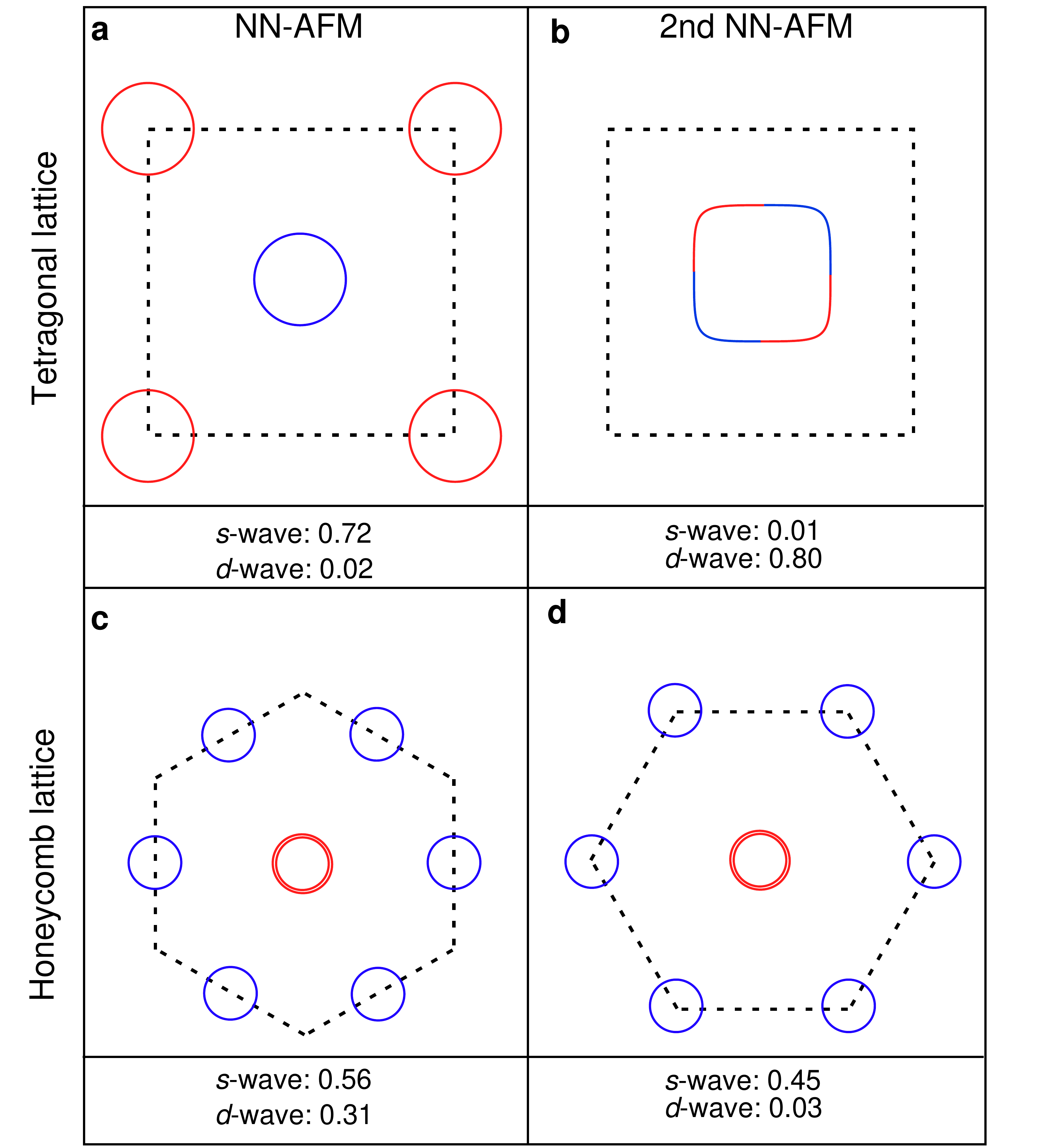}
\caption{  {\bf Predictions of possible collaborative Fermi surface topologies and AF exchange interactions that can result in undiscovered high-$T_c$ superconductors.}
 {\bf a,}  $s$-wave pairing in tetragonal lattice  with the NN AF exchange interactions.
 {\bf b,} $d$-wave pairing  in tetragonal lattice with the 2$_{nd}$ NN AF exchange interactions.
 {\bf c,} $s$-wave in honeycomb lattice with the NN exchange coupling.
 {\bf d,} $s$-wave in honeycomb lattice with the 2$_{nd}$ NN exchange coupling. 
 The numbers indicate the overlap strength of the corresponding reciprocal symmetry forms  on Fermi surfaces. The red and blue colors indicate the sign change of superconducting order parameters on Fermi surfaces. \label{fig5}   }
\end{figure*}

\end{document}

\draft
\title{Supplementary  Materials}
%\author{Xu}
%\affiliation{Beijing National Laboratory for Condensed Matter
%Physics, Institute of Physics, Chinese Academy of Sciences, Beijing
%100080, China}

%\begin{abstract}
%
%\end{abstract}

\maketitle

We briefly discuss  reciprocal pairing form factors provided by antiferromagnetic exchange interactions.    Considering a magnetic exchange coupling between two electrons at two different sites,  we have
\begin{eqnarray}
J_{ij}\vec S_i\cdot \vec S_j=\frac{1}{4}\sum_{\sigma}[2J_{ij} c^+_{i\sigma}c^+_{j\sigma} c_{j\bar{\sigma}}c_{i\bar{\sigma}}+J_{ij} c^+_{i\sigma}c^+_{j\sigma} c_{j\sigma}c_{i\sigma}-J_{ij}c^+_{i\sigma}c^+_{j\bar{\sigma}} c_{j\bar{\sigma}}c_{i\sigma}]
\label{ex}
\end{eqnarray}
where $\bar{\sigma}$ labels the opposite spin direction of $\sigma$. Given an antiferromagnetic exchange coupling $J_{ij}>0$, the decoupling of the first two terms in Eq.~\ref{ex}  in  pairing channel leads to triplet pairing and costs energy.  The last term in Eq.~\ref{ex}  gives singlet pair and saves energy. Defining $\Delta_{ij}=<c^+_{i\sigma}c^+_{j\bar{\sigma}}>$,  we obtain that the energy saved from magnetic exchange coupling is given by
\begin{eqnarray}
<J_{ij}\vec S_i\cdot \vec S_j>=-\frac{1}{2}J_{ij}|\Delta_{ij}|^2
\end{eqnarray}
 In a uniform superconducting state, $\Delta_{ij}$ should be a function of $\vec r_i-\vec r_j$.  Therefore, we can define $\Delta_{\vec k}=\frac{1}{N}\sum_{<ij>}e^{i\vec k \cdot (\vec r_i-\vec r_j)}\Delta_{ij}=<c^+_{\sigma}(\vec k)c^+_{\bar\sigma}(-\vec k)>$ , where $N$ is the total number of $<ij>$ links and $c^+_{\sigma}(\vec k)$ is electron creation operators in momentum space.

First, we consider cases in a tetragonal lattice  and define $\vec x$, $\vec y$ as the unit vectors of the lattice.
\begin{itemize}
\item  $s$-wave pairing by NN AFM in a tetragonal lattice:  \\
$\Delta_{ii\pm\vec x}=\Delta_{ii\pm\vec y}=\Delta_0$, $ \Delta_{\vec k}=\frac{\Delta_0}{4}(e^{ik_x}+e^{-ik_x}+e^{ik_y}+e^{-ik_y})=\frac{\Delta_0}{2}(cosk_x+cosk_y)$.
\item  $d$-wave pairing by NN AFM in a tetragonal lattice:  \\
$\Delta_{ii\pm\vec x}=-\Delta_{ii\pm\vec y}=\Delta_0 $ , $\Delta_{\vec k}=\frac{\Delta_0}{4}(e^{ik_x}+e^{-ik_x}-e^{ik_y}-e^{-ik_y})=\frac{\Delta_0}{2}(cosk_x-cosk_y)$.
\item  $s$-wave pairing by 2$_{nd}$ NN AFM in a tetragonal lattice:  \\
$\Delta_{ii\pm(\vec x\pm\vec y)}=\Delta_0$, $ \Delta_{\vec k}=\frac{\Delta_0}{4}(e^{ik_x+ik_y}+e^{-ik_x-ik_y}+e^{ik_x-ik_y}+e^{ik_y-ik_x})=\Delta_0cosk_xcosk_y$.
\item  $d$-wave pairing by 2$_{nd}$  NN AFM in a tetragonal lattice:  \\
$\Delta_{ii\pm(\vec x+\vec y)}=-\Delta_{ii\pm(\vec x-\vec y)}=\Delta_0 $ , $\Delta_{\vec k}=\frac{\Delta_0}{4}(e^{ik_x+ik_y}+e^{-ik_x-ik_y}-e^{ik_x-ik_y}-e^{ik_y-ik_x})=\Delta_0 sink_xsink_y$.
\item  $s$-wave pairing by 3$_{rd}$ NN AFM in a tetragonal lattice:  \\
$\Delta_{ii\pm2\vec x}=\Delta_{ii\pm2\vec y}=\Delta_0$, $ \Delta_{\vec k}=\frac{\Delta_0}{4}(e^{i2k_x}+e^{-i2k_x}+e^{i2k_y}+e^{-i2k_y})=\frac{\Delta_0}{2}(cos2k_x+cos2k_y)$.
\item  $d$-wave pairing by 3$_{rd}$ NN AFM in a tetragonal lattice:  \\
$\Delta_{ii\pm2\vec x}=-\Delta_{ii\pm2\vec y}=\Delta_0 $ , $\Delta_{\vec k}=\frac{\Delta_0}{4}(e^{i2k_x}+e^{-i2k_x}-e^{i2k_y}-e^{-i2k_y})=\frac{\Delta_0}{2}(cos2k_x-cos2k_y)$.
\end{itemize}
Second, we consider a standard triangle lattice with the two unit vectors  $\vec e_1=(1,0),\vec e_2=(\frac{1}{2},\frac{\sqrt{3}}{2})$.
\begin{itemize}
\item  $s$-wave pairing by NN AFM in a triangle lattice:  \\
$\Delta_{ii\pm\vec e_1}=\Delta_{ii\pm\vec e_2}=\Delta_{ii\pm(\vec e_1-\vec e_2)}=\Delta_0$,\\
 $ \Delta_{\vec k}=\frac{\Delta_0}{6}(e^{i\vec k\cdot \vec e_1}+e^{-i\vec k\cdot \vec e_1}+e^{i\vec k\cdot \vec e_2}+e^{-i\vec k\cdot \vec e_2}+
e^{i\vec k\cdot (\vec e_1-\vec e_2)}+e^{-i\vec k\cdot(\vec e_1-\vec e_2)})=\frac{\Delta_0}{3}(cosk_x+2cos\frac{k_x}{2}cos\frac{\sqrt{3}}{2}k_y)$.
\item  $d$$\pm$i$d$-wave pairing by NN AFM in a triangle lattice:  \\
$\Delta_{ii\pm\vec e_1}=e^{\pm i\frac{2\pi}{3}}\Delta_{ii\pm\vec e_2}=e^{\pm i\frac{4\pi}{3}}\Delta_{ii\pm(\vec e_1-\vec e_2)}=\Delta_0$,\\
$ \Delta_{\vec k}^\pm=\frac{\Delta_0}{6}(e^{i\vec k\cdot \vec e_1}+e^{-i\vec k\cdot \vec e_1}+e^{\pm i \frac{2\pi}{3}}(e^{i\vec k\cdot \vec e_2}+e^{-i\vec k\cdot \vec e_2})+e^{\pm i\frac{4\pi}{3}}
(e^{i\vec k\cdot (\vec e_1-\vec e_2)}+e^{-i\vec k\cdot(\vec e_1-\vec e_2)}))=\frac{\Delta_0}{3}(cosk_x-cos\frac{k_x}{2}cos\frac{\sqrt{3}}{2}k_y\pm i\sqrt{3}sin\frac{k_x}{2}sin\frac{\sqrt{3}}{2}k_y)$.
\end{itemize}
Finally, we consider  a honeycomb lattice where the two unit vectors are given by $\vec e_1=(\frac{\sqrt{3}}{2},\frac{1}{2}), \vec e_2=(\frac{\sqrt{3}}{2},-\frac{1}{2})$. For convenience, we define $\vec e_0=(-\frac{1}{\sqrt{3}},0)$.
\begin{itemize}
\item  $s$-wave pairing by NN AFM in a honeycomb lattice:  \\
$\Delta_{ii+\vec e_0}=\Delta_{ii+\vec e_0+\vec e_1}=\Delta_{ii+(\vec e_1+\vec e_2)}=\Delta_0$,\\
 $ \Delta_{\vec k}=\frac{\Delta_0}{3}(e^{i\vec k\cdot \vec e_0}+e^{i\vec k\cdot (\vec e_0+\vec e_2)}+
e^{i\vec k\cdot (\vec e_0+\vec e_2)})=\frac{\Delta_0}{3}e^{-i\frac{1}{\sqrt{3}}k_x}(1+2cos(\frac{k_y}{2})e^{i\frac{\sqrt{3}}{2}k_x})$.
\item  $d$$\pm$i$d$-wave pairing by NN AFM in a honeycomb lattice:  \\
$\Delta_{ii＋\vec e_0}=e^{\pm i\frac{4\pi}{3}}\Delta_{ii＋\vec e_0+\vec e_1}=e^{\pm i\frac{2\pi}{3}}\Delta_{ii+(\vec e_0+\vec e_2)}=\Delta_0$,\\
$ \Delta_{\vec k}=\frac{\Delta_0}{3}(e^{i\vec k\cdot \vec e_0}+e^{\pm i\frac{2\pi}{3}}e^{i\vec k\cdot (\vec e_0+\vec e_2)}+e^{\pm i\frac{4\pi}{3}}
e^{i\vec k\cdot (\vec e_0+\vec e_2)})=\frac{\Delta_0}{3}e^{-i\frac{1}{\sqrt{3}}k_x}(1+2cos(\frac{k_y}{2}\pm\frac{2\pi}{3})e^{i\frac{\sqrt{3}}{2}k_x})$.
\item  $s$-wave pairing by 2$_{nd}$ NN AFM in a honeycomb lattice:  \\
$\Delta_{ii\pm\vec e_1}=\Delta_{ii\pm\vec e_2}=\Delta_{ii\pm(\vec e_1-\vec e_2)}=\Delta_0$,\\
 $ \Delta_{\vec k}=\frac{\Delta_0}{6}(e^{i\vec k\cdot \vec e_1}+e^{-i\vec k\cdot \vec e_1}+e^{i\vec k\cdot \vec e_2}+e^{-i\vec k\cdot \vec e_2}+
e^{i\vec k\cdot (\vec e_1-\vec e_2)}+e^{-i\vec k\cdot(\vec e_1-\vec e_2)})=\frac{\Delta_0}{3}(cosk_y+2cos\frac{k_y}{2}cos\frac{\sqrt{3}}{2}k_x)$.
\item  $d$$\pm$i$d$-wave pairing by  2$_{nd}$ NN honeycomb in a honeycomb lattice:  \\
$\Delta_{ii\pm\vec e_1}=e^{\pm i\frac{2\pi}{3}}\Delta_{ii\pm\vec e_2}=e^{\pm i\frac{4\pi}{3}}\Delta_{ii\pm(\vec e_1-\vec e_2)}=\Delta_0$,\\
$ \Delta_{\vec k}^\pm=\frac{\Delta_0}{6}(e^{i\vec k\cdot \vec e_1}+e^{-i\vec k\cdot \vec e_1}+e^{\pm i \frac{2\pi}{3}}(e^{i\vec k\cdot \vec e_2}+e^{-i\vec k\cdot \vec e_2})+e^{\pm i\frac{4\pi}{3}}
(e^{i\vec k\cdot (\vec e_1-\vec e_2)}+e^{-i\vec k\cdot(\vec e_1-\vec e_2)}))=\frac{\Delta_0}{3}(cosk_y-cos\frac{k_x}{2}cos\frac{\sqrt{3}}{2}k_x\pm i\sqrt{3}sin\frac{k_y}{2}sin\frac{\sqrt{3}}{2}k_x)$.
\end{itemize}
We define the overlap between  reciprocal form factors $\Delta_{\vec k}$ and Fermi surfaces  as
\begin{eqnarray}
W=\int\int dk_xdk_y |\Delta_{\vec k}|^2\delta(\epsilon_{\vec{k}}-\mu)
\end{eqnarray}
To perform numerical calculations, we  evaluate the above formular as follows
\begin{eqnarray}
W=\frac{\int\int dk_xdk_y |\Delta_{\vec k}|^2 \Theta(\omega-|\epsilon_{\vec{k}}-\mu|)}{\int\int dk_xdk_y \Theta(\omega-|\epsilon_{\vec{k}}-\mu|)}
\end{eqnarray}
where $\omega$ is a small positive value that is much less than the band width and $\Theta(x)$ is the unit step funciton defined as $\Theta(x)=1(0)$ if $x>0(x\leq0)$. $W$ has very week dependence on $\omega$. For a multi-band system with N bands, we  evaluate $W_{\alpha}$ for each band and define the average weight as $ W=\frac{\sum_{\alpha}W_{\alpha}}{N}$.  This average weight is a good quantity to evaluate  approximately the overlap strength  in iron-based superconductors since the gap functions of all the bands are fitted to a single pairing form function as disucssed in this paper.